\documentclass[conference]{IEEEtran}
\IEEEoverridecommandlockouts
\usepackage{cite}
\usepackage{amsmath,amssymb,amsfonts}
\usepackage{algorithmic}
\usepackage{graphicx}
\usepackage{textcomp}
\usepackage{xcolor}
\def\BibTeX{{\rm B\kern-.05em{\sc i\kern-.025em b}\kern-.08em
    T\kern-.1667em\lower.7ex\hbox{E}\kern-.125emX}}
\begin{document}

\title{Deepfake Detection of Singing Voices With Whisper Encodings}

\author{\IEEEauthorblockN{Falguni Sharma}
\IEEEauthorblockA{\textit{Centre for Artificial Intelligence} \\
\textit{Banasthali Vidyapith}\\
Jaipur, India \\
fsharma0207@gmail.com}
\and
\IEEEauthorblockN{Priyanka Gupta}
\IEEEauthorblockA{\textit{Dept. of Communication and Computer Engineering} \\
\textit{The LNM Institute of Information Technology}\\
Jaipur, India \\
priyanka.gupta@lnmiit.ac.in}}

\maketitle

\begin{abstract}
The deepfake generation of singing vocals is a concerning issue for artists in the music industry. In this work, we propose a singing voice deepfake detection (SVDD) system, which uses noise-variant encodings of open-AI's Whisper model. As counter-intuitive as it may sound, even though the Whisper model is known to be noise-robust, the encodings are rich in non-speech information, and are noise-variant. This leads us to evaluate Whisper encodings as feature representations for the SVDD task. Therefore, in this work, the SVDD task is performed on vocals and mixtures, and the performance is evaluated in \%EER over varying Whisper model sizes and two classifiers- CNN and ResNet34, under different testing conditions.
\end{abstract}

\vspace{-0.1cm}
\section{Introduction}
\vspace{-0.1cm}
The growing improvements in generative models have considerably enhanced automatic human voice generation, with nearly natural-parity performances. However, such an artificial audio data generation has paved the way for forgeries, popularly known as \emph{deepfakes} \cite{wang2024asvspoof, gupta2024vulnerability}. Particularly in the music industry, unauthorized deepfake copies that closely resemble well-known vocalists are a growing source of concern for musicians. These reproductions pose a direct threat to the commercial worth and intellectual property rights of the original artists. Some of the singing voice synthesis models such as VISinger \cite{zhang2022visinger}, DiffSinger \cite{liu2022diffsinger}, MidiVoices \cite{byun2024midi}, and SinTechSVS \cite{zhao2024sintechsvs} have the capability to synthesize convincing levels of natural sounding vocals, thereby mimicking vocalists well. Therefore, the effective detection of deepfake-generated music calls for attention.
\par As opposed to speech spoof detection, singing voice deepfake detection (SVDD), presents a unique set of issues that are not observed in speech. For example, the pitch and duration of phonemes in a singing voice are significantly affected by the melody. Moreover, the wider range of timbre and voicing by artists in a musical context is not observed in speech. Additionally, extensive editing in singing voices is done to mix with musical instrumental accompaniments. These contrasts with speech pose the question of whether the countermeasures designed for speech can be directly applied to SVDD. In this context, research in Singing Voice Deepfake Detection (SVDD) has come up recently, with the proposition of the \emph{Singfake} dataset \cite{zang2024singfakesingingvoicedeepfake}. This was followed by the SVDD 2024 challenge, having CtrSVDD and WildSVDD tracks, with its evaluation labels yet to be released. Recently, SingGraph has been proposed as a state-of-the-art model for singing voice deepfake detection, integrating music and linguistic analysis with domain-specific augmentation techniques. Our work proposes a different approach, focusing on non-invariant features without using data augmentation techniques, to address the challenges in this domain.

\par We propose an end-to-end model for SVDD which utilizes the encoder representations from the pre-trained \textbf{W}eb-scale \textbf{S}upervised \textbf{P}retraining for \textbf{S}peech \textbf{R}ecognition (WSPR) model also referred to as \emph{Whisper} \cite{radford2023robust}.  Whisper is incredibly resilient to real-world background sounds, like music. However counter-intuitively, unlike the other Automatic Speech Recognition (ASR) systems, its audio representation is not noise-invariant \cite{gong2023whisper}. Therefore, we use these noise-variant encodings from the encoder output of the Whisper model, for the SVDD task. In particular, this paper has the following contributions:
\begin{itemize} 
    \item End-to-end Whisper encoding-based SVDD system is proposed.
    \item Noise-variance in encodings is used as a discriminative cue to classify the bonafide and deepfake voices, in pure vocals, as well as in the case of mixtures.
    \item Performance evaluation is done on four variants (W(Tiny), W(Small), W(Base), and W(Medium)) of Whisper.
    \item Experiments on vocals and mixtures are also done to investigate the effect of instrumental accompaniments in the songs. 
    \item Experimental analysis w.r.t. various testing conditions is done to observe the effect of seen/unseen singers, different communication codecs, languages, and musical contexts. 
\end{itemize}
\begin{figure*}[!h]
    \centering
\includegraphics[width=\linewidth]{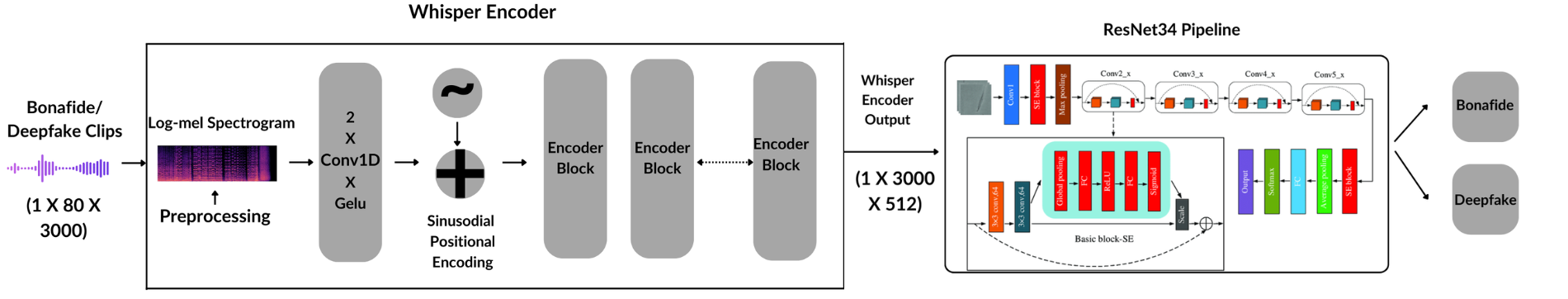}
    \caption{The proposed end-to-end SVDD system using Whisper Encoder with ResNet34}
    \vspace{-0.2cm}
    \label{fig:Architecture}
\end{figure*}

\vspace{-0.2cm}
\section{Proposed Work}
\vspace{-0.2cm}
\label{sec:proposed}
\subsection{Whisper Encoder}
\vspace{-0.2cm}
Whisper is an extensively trained ASR model, with training done on 680K hours of data collected from the Internet with diverse environments and recording setups. It is based on encoder-decoder Transformer architecture, primarily given by \cite{vaswani2023attentionneed} and it is trained on a substantial and varied supervised dataset and focuses on zero-shot transfer. In this work, we propose a SVDD system, which uses Whisper encodings along with ResNet34 in the pipeline, as shown in Figure \ref{fig:Architecture}. 

The encoder of the Whisper model from which the encodings are extracted takes input audio, which is pre-processed to mel-spectrogram. The corresponding mel-spectrogram is processed through two 1-D convolutional layers to extract features. Additionally, sinusoidal positional embeddings are then added for temporal context. The variant of the Whisper models are with respect to the size of its encoder. In particular, tiny model incorporates 4 blocks, the base model uses 6, the small model employs 12, and the medium model utilizes 24 blocks and according to that it gives a vector output of fixed dimensions in its last hidden state. Each block consists of a multi-head self-attention mechanism,
\begin{equation}
\text{Attention}(Q, K, V) = \text{softmax}\left(\frac{Q K^T}{\sqrt{d_k}}\right) V
\end{equation} 
 where $Q \in \mathbb{R}^{T \times d_k}$, keys $K \in \mathbb{R}^{T \times d_k}$, and values $V \in \mathbb{R}^{T \times d_v}$, where T is the sequence length, and $d_k$ and $d_v$ are the hidden dimensionality for queries/keys and values respectively. Following this is a position-wise fully connected feed-forward network. The output of each sub-layer is LayerNorm(x + \emph{Sublayer}(\emph{x})), where \emph{Sublayer}(\emph{x}) is the function implemented by the sub-layer itself.
    \vspace{-0.1cm}
\subsection{Noise-Variant Encodings from Noise Robust Whisper}
\vspace{-0.1cm}
There have been various studies that focus on noise-invariant representations of ASR \cite{serdyuk2016invariant, liang2018learning, zhu2022noise}. Researchers often specify noise invariance as an explicit inductive bias for robust ASR since it is widely accepted that a robust ASR model's representation should be noise-invariant \cite{van2009unsupervised, serdyuk2016invariant, liang2018learning, zhu2022noise}. For Whisper however, a counter-intuitive finding is that Whisper's audio representation is \emph{not noise-invariant}; rather, it encodes rich information about non-speech background sounds. This indicates that the Whisper model encodes the type of noise and subsequently detects speech conditioned on it, instead of learning a noise-invariant representation \cite{gong2023whisper}. Since the training data of Whisper is very large (680K hours), it consists of diverse settings of environments, languages, and speakers, and has noisy labels. This differentiates the Whisper model from the other ASR models. As opposed to learning a representation that is independent of noise, it \emph{encodes} the background sound first and then transcribes text based on the type of noise \cite{gong2023whisper}.
\par
\begin{figure}[!h]
    \centering
\includegraphics[width=\linewidth]{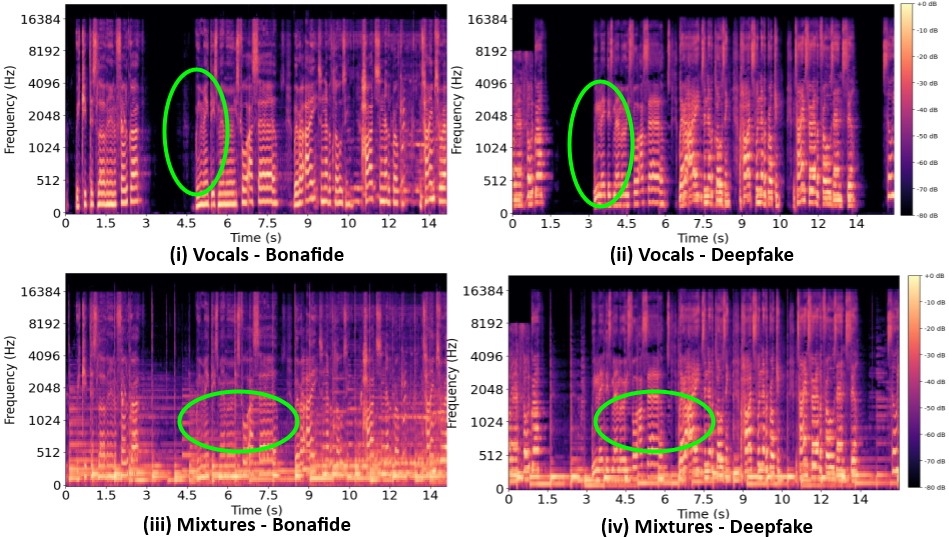}
         \caption{Bonafide vs. Deepfake Singing Voices}
    \label{fig:Spectrogram}
\end{figure}

 We leverage Whisper's \emph{noise-conditioned} encodings to capture the noise introduced in deepfakes and use it as the discriminative feature/cue to distinguish between bonafide and deepfake singing voices. In this work, the Singfake dataset has been used which is divided into two scenarios - vocals (containing only the singer's voice), and mixtures (the singer's voice is accompanied by background music). In context with this, Figure \ref{fig:Spectrogram} shows the spectrographic comparison of bonafide vs. deepfake singing voices for both vocals and mixtures.
In particular, as indicated by the regions marked in green circles in Figure \ref{fig:Spectrogram}, in the case of vocals, the end of a non-vocal region (silent region), has a gradual fading in bonafide utterance (Figure \ref{fig:Spectrogram} (i)). However, in the deepfaked singing voice, a very distinct and sharp transition from the silence region to the singing region can be observed (Figure \ref{fig:Spectrogram} (ii)). Furthermore, in the case of mixtures, the formants of the singer's voice dissolve into the background instrument frequencies, as observed by the region marked in green in Figure \ref{fig:Spectrogram} (iii). However, in the case of deepfaked mixture, the formants appear very distinct from the background instrumentals, as observed by the region marked in Figure \ref{fig:Spectrogram} (iv). 
\vspace{-0.2cm}
\section{Experimental Setup}
\vspace{-0.2cm}
\label{sec:setup}
\subsection{Dataset}
\vspace{-0.2cm}
For this study, the Singfake dataset \cite{zang2024singfakesingingvoicedeepfake} is used. It offers $28.93$ hours of bonafide (real songs) and $29.40$ hours of deepfake generated songs, with partitioning details as shown in Table \ref{tab:Dataset Statistic for each split}. Furthermore, the Singfake dataset has two subsets- vocals, and mixtures.
\begin{table}[!h]
    \centering
    \caption{Singfake Data Partitions}
    \label{tab:Dataset Statistic for each split}
    \resizebox{0.8\columnwidth}{!}{%
    \begin{tabular}{|l|l|l|}
    \hline
        Partition & Description & Bonafide/Deepfake \\ \hline
        Train & General Training Set &  5251 / 4519 \\ \hline
        Validation & General Validation Set & 1089 / 543  \\ \hline
        T01 & Seen Singers, Unseen Songs & 370 / 1208  \\ \hline
        T02 & Unseen Singers, Unseen Songs &  1685 / 1006 \\ \hline
        T03 & T02 over 4 communication codecs & 6740 / 4024 \\ \hline
        T04 & Unseen Languages/musical Contexts & 353 / 166 \\ \hline
    \end{tabular}%
    }
\end{table}
 The vocals subset consists of separated singing vocals from the background instrument accompaniments. For vocal separation, Demucs \cite{défossez2019demucsdeepextractormusic} was used, and each song was separated into $6-8$ clips of an average of $13.75$ seconds duration, using  Voice Activity Detection (VAD) from the Pyannote library \cite{9604948, bredin2020pyannote}.
Furthermore, the mixture subset consists of songs with instrumental accompaniments. Similar to the vocals, the songs were clipped into segments based on the timestamps obtained by VAD. This resulted in the mixture clips similar to vocal clips with background accommodations. 
\vspace{-0.2cm}
\subsection{Classifier and Performance Metrics} 
\vspace{-0.1cm}
Two classifiers are utilized in this study: Convolutional Neural Network (CNN) and ResNet34. The CNN architecture in this work consists of two convolutional layers, followed by two max-pooling layers and a fully connected layer. The kernel size is kept as $5$ and the activation function used is ReLU. The ResNet34 model consists of $34$ residual networks comprising a $7\times7$ convolution layer, subsequently one max pooling layer and four stages of residual blocks with $3\times3$ convolutional filters and identity shortcuts, and a final global average pooling and fully connected layer. Both the classifiers use the Adam optimizer with a learning rate of $0.001$ and a batch size of $32$. Furthermore, the Equal Error Rate (EER) is used as the performance metric.
\vspace{-0.2cm}
\subsection{Baseline}
\vspace{-0.1cm}
We have considered  Wav2vec2+AASIST as the baseline in this work \cite{zang2024singfakesingingvoicedeepfake}. It is the best-performing system on the Singfake dataset so far. It should be noted that in this baseline work \cite{zang2024singfakesingingvoicedeepfake}, the RawBoost data augmentation module has been removed for fair comparisons between methods.

\vspace{-0.1cm}
 \section{Experimental Results}
\vspace{-0.1cm}
\label{sec:results}
\subsection{Whisper Encodings vs. Standard Representations}
\label{subsec:results_encodings}
This subsection shows the experimental results of the proposed Whisper-encodings-based-SVDD system,  of varying sizes- tiny, base, small, and medium, denoted as W(Tiny), W(Base), W(Small) and W(Med.), respectively. To observe the effectiveness of the proposed method, additional experiments w.r.t. \emph{non-Whisper} standard feature sets MFCC, and CQCC were performed. The experimental results corresponding to LFCC are taken from \cite{zang2024singfakesingingvoicedeepfake}.

\subsubsection{Vocals}
\label{subsubsec:encode_vocals}
\begin{figure}[!h]
    \centering   
    \includegraphics[width=\linewidth]{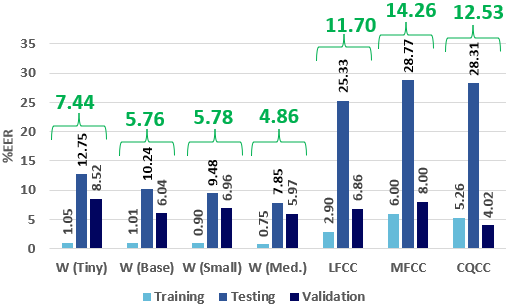}
    \caption{\%EER for various splits on Singfake vocals across different representations, with ResNet34 as the classifier}
\label{fig:Blue_ResNet_with_averages}
\end{figure}

\begin{figure}[!h]
    \centering
\includegraphics[width=\linewidth]{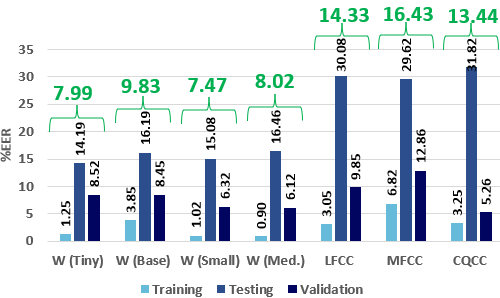}
    \caption{\%EER for various splits on Singfake vocals across different representations, with CNN as the classifier}
\label{fig:Blue_CNN_with_averages}
\end{figure}

 Figure \ref{fig:Blue_ResNet_with_averages} and Figure \ref{fig:Blue_CNN_with_averages} show the training, testing (average of the performances of all the 4 test cases), and validation performances on classifiers ResNet34 and CNN, respectively, for the case when only vocals from the dataset are considered.
It can be observed that for \emph{both} the classifiers, the proposed approach of extracting Whisper encodings significantly outperforms \emph{all} the standard feature sets. In particular, with ResNet34 (as shown in Figure \ref{fig:Blue_ResNet_with_averages}), encodings from the medium-sized Whisper model (denoted as W (Med.)) achieve an average EER of $4.86\%$. 
\subsubsection{Mixtures}
\begin{figure}[!h]
    \centering
\includegraphics[width=0.92\linewidth]{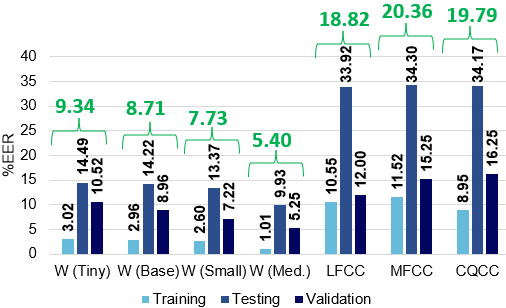}
    \caption{\%EER for various splits on Singfake mixtures across different representations, with ResNet34 as the classifier}
\label{fig:Blue_Mixture_ResNet_avg}
\end{figure}

\begin{figure}[!h]
    \centering
\includegraphics[width=0.92\linewidth]{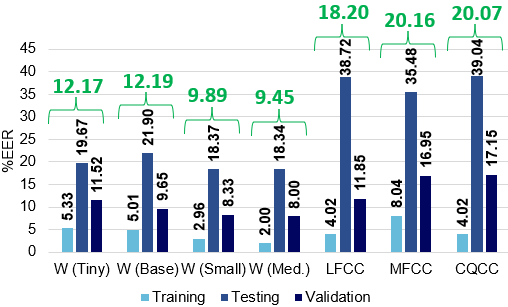}
    \caption{\%EER for various splits on Singfake mixtures across different representations, with CNN as the classifier}
\label{fig:Blue_Mixture_CNN_avg}
\end{figure}

For the case of mixtures (i.e., singing voices that have background instrumental accompaniments), Figure \ref{fig:Blue_Mixture_ResNet_avg} and Figure \ref{fig:Blue_Mixture_CNN_avg}, show the comparative performances on the training, average testing, and validation sets, using classifiers ResNet34 and CNN, respectively.
It can be observed that for mixtures as well, for \emph{both} the classifiers, the encoding from the Whisper-medium model significantly outperforms \emph{all} the standard feature sets, with an average EER of $9.45\%$. It should also be noted that all the Whisper encoding-based systems \emph{significantly outperform} the non-Whisper respresentation-based SVDD systems.  
\subsection{Analysis on Testing Conditions}
We now analyze our findings for different test cases (T01 to T04). Given that in the previous subsection \ref{subsec:results_encodings}, the ResNet34 classifier showed the best performance on both vocals and mixtures, we now consider ResNet34 at the classifier end for the rest of the experiments in this paper. 
\subsubsection{Vocals}
\begin{figure}[!h]
    \centering
    \includegraphics[scale=0.4]{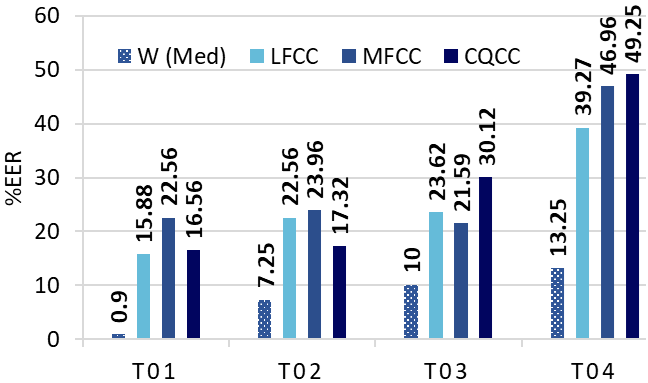}
    \caption{Testing Conditions Comparison on Vocals}
    \label{fig:testing_vocals}
\end{figure}

Figure \ref{fig:testing_vocals} shows the experimental results obtained on various testing conditions (T01 to To4) such as seen/unseen singers \& languages, communication codecs, and musical contexts.  It is observed that for \emph{all} the systems, T01 is the easiest, and T04 is the hardest to detect. In particular, on the T01 case (seen singers, but unseen songs), the best performance of $1.09\%$ EER is achieved by the proposed system based on W(Med.) with ResNet34 as the classifier.
\subsubsection{Mixtures}
\begin{figure}[!h]
    \centering
    \includegraphics[scale=0.4]{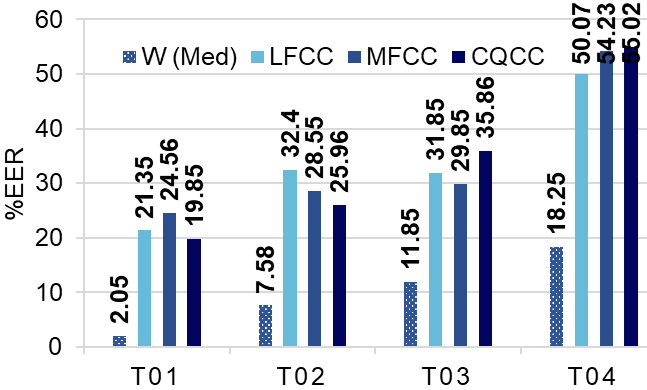}
    \caption{Testing Conditions Comparison on Mixtures}
    \label{fig:testing_mixtures}
\end{figure}

It can be observed from Figure \ref{fig:testing_mixtures}, that the performances on mixtures follow the same trend as for the case of vocals. Additionally, for all the values of the EERs for all the representations, the vocals are found to be easier to detect than deepfakes in the mixture. Overall, the proposed Whisper encodings-based system is observed to outperform \emph{all} the remaining systems, with remarkable differences in EER values, for \emph{both} vocals and mixtures. However, with unseen language and musical style, T04 is the most difficult to detect, indicating the limitation of the models. 
\subsection{Comparison with Existing Systems}
We compare the performance of our proposed model- Whisper(Med.) with ResNet34,  against the best-performing model (Wav2vec2 + AASIST) on Singfake \cite{zang2024singfakesingingvoicedeepfake}. It is to be mentioned that the Singfake paper did not provide results for the validation dataset. The comparison of both the vocals and the mixture highlights that the Whisper(medium) based model consistently performed better across all training and testing conditions, especially in T04, with an absolute difference of $28.94\%$ and $24.52$, in vocals and mixtures, respectively.


\begin{figure}[!h]
    \centering
    \includegraphics[width=0.98\linewidth]{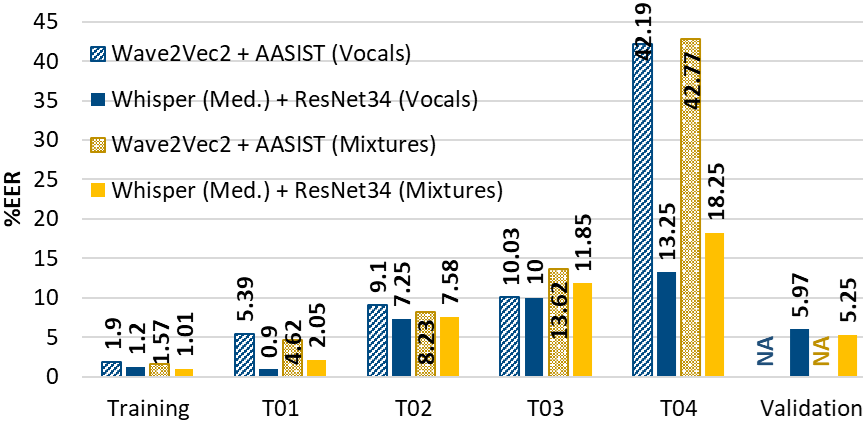}
    \caption{Comparison of the proposed system with the baseline}
    \label{fig:compare_baselines}
\end{figure}
\section{Conclusions and Future Works}
\label{sec:conclude}
This work proposed noise-variant encodings from the Whisper model as representations to be used for SVDD. It was observed that the encodings from the Whisper (medium) model with ResNet34 classifier achieved the best performances throughout all the testing scenarios of the Singfake dataset. Our findings showed that amongst the \emph{non-whisper} representations, LFCC showed the best performance. 

However, our system's robustness under T04 conditions, which involve unseen languages, remains a limitation. Although our results surpass those of the baseline paper, there is room for improvement in handling such diverse linguistic conditions. In the future, the effectiveness of the proposed SVDD system can also be improved by incorporating data augmentation.

\bibliographystyle{IEEEtran}

\end{document}